\documentclass[aps,prd,twocolumn,
preprintnumbers,
showpacs,floatfix]{revtex4}

\usepackage{bm}
\usepackage{epsfig}
\usepackage{graphics}
\usepackage{psfrag}
\usepackage{amsmath}
\usepackage{textcomp}

\newcommand{\eeq}{\end{equation}}
\newcommand{\beq}{\begin{equation}}
\newcommand{\ba}{\begin{array}}
\newcommand{\ea}{\end{array}}
\newcommand{\bea}{\begin{eqnarray}}
\newcommand{\eea}{\end{eqnarray}}
\newcommand{\vev}[1]{\langle #1\rangle}

\newcommand{\veps}{\varepsilon}

\newcommand{\si}{\sigma}

\newcommand{\lsim}{\mbox{\raisebox{-.9ex}
{~$\stackrel{\mbox{$<$}}{\sim}$~}}}
\newcommand{\gsim}{\mbox{\raisebox{-.9ex}
{~$\stackrel{\mbox{$>$}}{\sim}$~}}}

\newcommand{\pb}{\bar{\phi}}
\newcommand{\Hb}{\bar{H}}
\newcommand{\ka}{\kappa}
\newcommand{\la}{\lambda}

\newcommand{\ten}[1]{\,\cdot 10^{#1}}
\newcommand{\units}[1]{\,\text{#1}}

\DeclareMathOperator{\tr}{tr}

\begin{document}

\preprint{UT-STPD-4/07}

\title{Standard-smooth hybrid inflation}

\author{George Lazarides}
\email{lazaride@eng.auth.gr}
\author{Achilleas Vamvasakis}
\email{avamvasa@gen.auth.gr}
\affiliation{Physics Division, School of
Technology, Aristotle University of
Thessaloniki, Thessaloniki 54124, Greece}

\date{\today}

\begin{abstract}
We consider the extended supersymmetric
Pati-Salam model which, for $\mu>0$ and universal
boundary conditions, succeeds to yield
experimentally acceptable $b$-quark masses by
moderately violating Yukawa unification. It is
known that this model can lead to new shifted or
new smooth hybrid inflation. We show that a
successful two-stage inflationary scenario can
be realized within this model based only on
renormalizable superpotential interactions. The
cosmological scales exit the horizon during the
first stage of inflation, which is of the
standard hybrid type and takes place along the
trivial flat direction with the inflaton driven
by radiative corrections. Spectral indices
compatible with the recent data can be achieved
in global supersymmetry or minimal supergravity
by restricting the number of e-foldings of our
present horizon during the
first inflationary stage. The additional
e-foldings needed for solving the horizon and
flatness problems are naturally provided by a
second stage of inflation, which occurs mainly
along the built-in new smooth hybrid inflationary
path appearing right after the destabilization of
the trivial flat direction at its critical point.
Monopoles are formed at the end of the first
stage of inflation and are, subsequently, diluted
by the second stage of inflation to become
utterly negligible in the present universe for
almost all (for all) the allowed values of the
parameters in the case of global supersymmetry
(minimal supergravity).
\end{abstract}

\maketitle

\section{Introduction}
\label{sec:intro}

\par
It is well known \cite{smooth1} that the standard
supersymmetric (SUSY) realization \cite{cop,rc}
of hybrid inflation \cite{linde} in the context
of grand unified theories (GUTs) leads, at the
end of inflation, to a copious production of
topological defects such as cosmic strings
\cite{string}, magnetic monopoles
\cite{monopole}, or domain walls \cite{domain} if
these defects are predicted by the underlying
symmetry breaking. In the case of magnetic
monopoles or domain walls, this causes a
cosmological catastrophe. The simplest GUT gauge
group whose breaking to the standard model (SM)
gauge group $G_{\rm SM}$ predicts the existence
of topologically stable magnetic monopoles is
the Pati-Salam (PS) group $G_{\rm PS}=
{\rm SU}(4)_c\times{\rm SU}(2)_{\rm L}\times
{\rm SU}(2)_{\rm R}$ \cite{pati}. (Note that
the PS monopoles carry \cite{magg} two units of
Dirac magnetic charge.) So, applying the standard
realization of hybrid inflation within the SUSY
PS GUT model, we encounter a cosmologically
disastrous overproduction of magnetic monopoles
at the end of inflation, where the GUT gauge
symmetry $G_{\rm PS}$ breaks spontaneously to
$G_{\rm SM}$.

\par
Possible ways out of this difficulty are provided
by the shifted \cite{shift} or smooth
\cite{smooth1,smooth2}
variants of SUSY hybrid inflation, which, in
their conventional realization, utilize
non-renormalizable superpotential terms (for a
review, see Ref.~\cite{talks}). In these
inflationary scenarios, the GUT gauge symmetry
$G_{\rm PS}$ is broken to $G_{\rm SM}$ already
during inflation and, thus, no magnetic monopoles
are produced at the termination of inflation.

\par
It has been shown \cite{nshift,nsmooth} that
hybrid inflation of both the shifted and smooth
type can be implemented within an extended SUSY
PS model without the need of non-renormalizable
superpotential interactions. It is very
interesting to note that this extended SUSY PS
model was initially constructed \cite{quasi}
(see also Ref.~\cite{quasitalks}) for solving a
very different problem. In SUSY models with exact
Yukawa unification \cite{als}, such as the
simplest SUSY PS model (see Ref.~\cite{hw}), and
universal boundary conditions, the $b$-quark mass
comes out \cite{hall} unacceptably large for
$\mu>0$. Therefore, Yukawa unification must be
moderately violated so that, for $\mu>0$, the
predicted $b$-quark mass resides within the
experimentally allowed range even with universal
boundary conditions. This requirement forces us
to extend the superfield content of the SUSY PS
model by including, among other superfields, an
extra pair of ${\rm SU}(4)_c$ non-singlet
${\rm SU}(2)_{\rm L}$ doublets, which naturally
develop \cite{wetterich} subdominant vacuum
expectation values (VEVs) and mix with the main
electroweak doublets of the model leading to a
moderate violation of Yukawa unification. It is
quite remarkable that the resulting extended SUSY
PS model can automatically and naturally lead
\cite{nshift,nsmooth} to a new version of shifted
and smooth hybrid inflation based solely on
renormalizable superpotential terms. As in the
conventional realization of shifted and smooth
hybrid inflation, the GUT gauge group
$G_{\rm PS}$ is broken to $G_{\rm SM}$ already
during inflation in the new shifted \cite{nshift}
and new smooth \cite{nsmooth} hybrid inflation
scenario too and monopole production at the end
of inflation is avoided.

\par
Unfortunately, there is generally a tension
between the above mentioned well-motivated,
natural, and otherwise successful hybrid
inflationary models and the recent three-year
results \cite{wmap3} from the Wilkinson microwave
anisotropy probe satellite (WMAP3). Indeed, these
models, with the exception of the smooth
\cite{smooth1} and especially the new smooth
\cite{nsmooth} hybrid inflation model, predict
that, in global SUSY, the spectral index
$n_{\rm s}$ is very close to unity and with no
much running. Moreover, inclusion of supergravity
(SUGRA) corrections with canonical K\"{a}hler
potential yields \cite{senoguz}, in all cases,
$n_{\rm s}$'s which are very close to unity or
even exceed it. On the other hand, fitting the
WMAP3 data with the standard power-law
cosmological model with cold dark matter and a
cosmological constant ($\Lambda$CDM), one obtains
\cite{wmap3} $n_{\rm s}$'s clearly lower than
unity.

\par
One possible resolution of this inconsistency is
\cite{lofti,king,rehman} to use a non-minimal
K\"{a}hler potential with a convenient choice of
the sign of one of its terms. This generates
\cite{king,rehman,gpp} a negative mass term for
the inflaton. Consequently, the inflationary
potential acquires, in general, a local maximum
and minimum. Then, as the inflaton rolls from
this maximum down to smaller values, hybrid
inflation of the hilltop type \cite{lofti} can
occur. In this case, $n_{\rm s}$ can become
consistent with the WMAP3 measurements, but only
at the cost of a mild tuning \cite{gpp} of the
initial conditions. In any case, we must make
sure that the system is not trapped in the local
minimum of the inflationary potential, which can
easily happen for general initial conditions. In
such a case, no hybrid inflation would take
place. Note that, in the cases of smooth and new
smooth hybrid inflation, acceptable $n_{\rm s}$'s
can be obtained \cite{rehman,nsmooth} even
without the appearance of this local maximum and
minimum and, thus, the related complications can
be avoided.

\par
Another possibility \cite{mhin} for reducing the
spectral index predicted by the hybrid
inflationary models is based on the observation
that, in these models, $n_{\rm s}$ generally
decreases with the number of e-foldings suffered
by our present horizon scale during hybrid
inflation. So, restricting this number of
e-foldings, we can achieve values of $n_{\rm s}$
which are compatible with the recent WMAP3 data
even with minimal K\"{a}hler potential. The
additional number of e-foldings required for
solving the horizon and flatness problems of
standard hot big bang cosmology can be provided
by a second stage of inflation which follows
hybrid inflation. In Ref.~\cite{mhin}, this
complementary inflation was taken to be of the
modular type \cite{modular} realized by a string
axion at an intermediate scale. Note, in passing,
that a restricted number of e-foldings during
hybrid inflation was previously used
\cite{yamaguchi} to achieve sufficient running
of the spectral index.

\par
In this paper, we reconsider the extended SUSY
PS model of Ref.~\cite{quasi} which solves the
$b$-quark mass problem and can lead to new
shifted \cite{nshift} or new smooth
\cite{nsmooth} hybrid inflation. We restrict
ourselves in the range of parameters of this
model that corresponds to the latter case. As
shown in Ref.~\cite{nsmooth}, the relevant scalar
potential possesses, in this case, a trivial
classically flat direction which is stable for
large values of the inflaton field. Along this
direction, the PS gauge group is unbroken. For
values of the inflaton field smaller than a
certain critical value, this flat direction is
destabilized giving its place to a classically
non-flat valley of minima along which new
smooth hybrid inflation can take place. The GUT
gauge group $G_{\rm PS}$ is broken to
$G_{\rm SM}$ in this valley.

\par
In Ref.~\cite{nsmooth}, we investigated the
possibility that all the cosmological scales exit
the horizon during new smooth hybrid inflation,
which is, thus, responsible for the observed
spectrum of primordial fluctuations. Here, we
will consider an alternative possibility. As
usual, the trivial flat direction acquires
\cite{rc} a logarithmic slope from one-loop
radiative corrections which are due to the SUSY
breaking caused by the non-vanishing potential
energy density on this direction. So, a version
of standard hybrid inflation can easily take
place as the system slowly rolls down the trivial
flat direction. We will assume here that the
cosmological scales exit the
horizon during this standard hybrid
inflation. Then, as in Ref.~\cite{mhin}, we can
easily achieve, in global SUSY, spectral indices
which are comfortably compatible with the data by
restricting the number of e-foldings suffered by
our present horizon scale during this
inflationary period. The additional number of
e-foldings required for solving the horizon and
flatness problems is naturally provided, in this
case, by a second stage of inflation consisting
of a relatively short intermediate inflationary
phase, which starts as soon as the system crosses
the critical point of the trivial flat direction,
followed by new smooth hybrid inflation. So, the
necessary complementary inflation is
automatically built in the model itself and we do
not have to invoke an {\it ad hoc} second stage
of inflation as in Ref.~\cite{mhin}. Moreover,
large reheat temperatures can, in principle, be
achieved after the second stage of inflation
since this stage is realized at a superheavy
scale. As a consequence, baryogenesis  via
thermal \cite{thermallepto} or
non-thermal \cite{nonthermallepto} leptogenesis
may work in this case in contrast to the model of
Ref.~\cite{mhin}, where the reheat temperature is
too low for the non-perturbative electroweak
sphalerons to operate. However, we should keep in
mind that, as in all SUSY theories, the presence
of flat directions (see e.g. Ref.~\cite{flat})
can \cite{quasithermal} naturally delay the
reheating and thermalization process which
follows the decay of the inflaton field. This
reduces the reheat temperature and, thus, may
severely constrain thermal leptogenesis.
Nevertheless, non-thermal leptogenesis remains a
viable alternative. Finally, the PS monopoles
which are formed at the end of the standard
hybrid stage of inflation can be adequately
diluted by the subsequent second stage of
inflation.

\par
The inclusion of SUGRA corrections with minimal
K\"{a}hler potential raises the spectral index,
which, however, remains acceptable for a wide
range of the model parameters. So, in this model,
there is no need to include non-minimal terms in
the K\"{a}hler potential and, consequently,
complications from the possible appearance of a
local maximum and minimum of the inflationary
potential are avoided.

\par
In Sec.~\ref{sec:hybsmooth}, we sketch the
salient features of the extended SUSY PS model
and show that it can easily lead to a successful
two-stage inflationary scenario. In
Sec.~\ref{sec:sugra}, we show that this scenario
remains viable even if SUGRA corrections with a
minimal K\"{a}hler potential are included and,
in Sec.~\ref{sec:gauge}, we discuss briefly
gauge unification. Finally, in
Sec.~\ref{sec:conclusions}, we summarize our
conclusions.

\section{Standard-smooth hybrid inflation in
global supersymmetry}
\label{sec:hybsmooth}

\par
We consider the extended SUSY PS model of
Ref.~\cite{quasi}. As mentioned, this model
admits a moderate violation of the asymptotic
Yukawa unification so that, for $\mu>0$, an
acceptable $b$-quark mass is obtained even with
universal boundary conditions. The breaking of
$G_{\rm PS}$ to $G_{\rm SM}$ is achieved by the
VEVs of the right handed neutrino type components
of a conjugate pair of Higgs superfields $H^c$
and $\bar{H}^c$ belonging to the $(\bar{4},1,2)$
and $(4,1,2)$ representations of $G_{\rm PS}$
respectively. The model also contains a gauge
singlet $S$ and a conjugate pair of superfields
$\phi$, $\bar{\phi}$ belonging to the (15,1,3)
representation of $G_{\rm PS}$. Note, in passing,
that almost all SUSY inflationary models involve
an {\it ad hoc} gauge singlet superfield such as
$S$ (an exception is the class of models of
Ref.~\cite{mssminf}). The superfield
$\phi$ acquires a VEV which breaks $G_{\rm PS}$
to $G_{\rm SM}\times{\rm U}(1)_{B-L}$. In
addition to $G_{\rm PS}$, the model possesses a
$Z_2$ matter parity symmetry and two global
${\rm U}(1)$ symmetries, which can effectively
arise \cite{laz1} from the rich discrete symmetry
groups encountered in many compactified string
theories (see e.g. Ref.~\cite{laz2}). For details
on the full field content and superpotential, the
charge assignments, and the phenomenological and
cosmological properties of this model, the reader
is referred to Refs.~\cite{quasi,shift} (see also
Ref.~\cite{quasitalks}).

\par
The superpotential terms which are relevant for
inflation are \cite{nsmooth}
\beq
\label{eq:superpotential}
W=\ka S(M^2-\phi^2)-\gamma S H^c\Hb^c+m\phi\pb
-\la\pb H^c\Hb^c,
\eeq
where $M$, $m$ are superheavy masses of the order
of the SUSY GUT scale $M_{\rm GUT}\simeq 2.86
\ten{16}\units{GeV}$ and $\ka$, $\gamma$, $\la$
are dimensionless coupling constants. All these
parameters are normalized so that they correspond
to the couplings between the SM singlet
components of the superfields. The mass
parameters $M$, $m$ and any two of the three
dimensionless parameters $\ka$, $\gamma$, $\la$
are made real and positive by appropriately
redefining the phases of the superfields. The
third dimensionless parameter, however, remains
in general complex. For definiteness, we choose
this parameter to be real and positive too as we
did in Ref.~\cite{nsmooth}.

\par
The F--term scalar potential obtained from the
superpotential $W$ in
Eq.~(\ref{eq:superpotential}) is given
\cite{nsmooth} by
\bea
\label{eq:SUSYpotential}
V&=&|\ka\,(M^2-\phi^2)-\gamma H^c\Hb^c|^2
\nonumber\\
& &+|m\pb-2\ka S\phi|^2+|m\phi-\la H^c\Hb^c|^2
\nonumber\\
& &+|\gamma S+\la\pb\,|^2\left(|H^c|^2+|\Hb^c|^2
\right),
\eea
where the complex scalar fields which belong to
the SM singlet components of the superfields are
denoted by the same symbol. In
Ref.~\cite{nsmooth}, it was shown that this
potential leads to a new version of smooth hybrid
inflation provided that
\beq
\label{eq:mu2}
\tilde{\mu}^2\equiv-M^2+\frac{m^2}{2\ka^2}>0
\eeq
and the parameter $\gamma$ is adequately small.
It was argued that, under these circumstances,
there exists a trivial classically flat direction
at $\phi=\pb=H^c=\Hb^c=0$ with
$V=V^0_{\rm tr}\equiv\ka^2 M^4$, which is a
valley of local minima for
\beq
|S|>S_c\equiv\sqrt{\frac{\ka}{\gamma}}\;M
\eeq
and becomes unstable for $|S|<S_c$, giving its
place to a classically non-flat valley of minima
along which new smooth hybrid inflation can take
place.

\par
We will now briefly summarize some of the main
results given in Ref.~\cite{nsmooth}. The SUSY
vacua of the potential in
Eq.~\eqref{eq:SUSYpotential} lie at
\begin{gather}
\phi=\frac{\gamma m}{2\ka\la}\left(
-1\pm\sqrt{1+\frac{4\ka^2\la^2M^2}
{\gamma^2m^2}}\,\right)\equiv\phi_{\pm},\\
\pb=S=0,\quad H^c\Hb^c=\frac{m}{\la}\,\phi.
\end{gather}
The vanishing of the D--terms yields
$\Hb^{c*}=e^{i\theta}H^c$, which implies that
there exist two distinct continua of SUSY
vacua:
\bea
\phi=\phi_{+},\quad\bar{H}^{c*}=H^c,\quad |H^c|=
\sqrt{\frac{m\phi_{+}}{\la}} \quad (\theta=0),
\label{eq:vacuum1}\\
\phi=\phi_{-},\quad\bar{H}^{c*}=-H^c,\quad |H^c|=
\sqrt{\frac{-m\phi_{-}}{\la}} \quad
(\theta=\pi)
\label{eq:vacuum2}
\eea
with $\pb=S=0$. One can show that the potential,
besides the trivial flat direction, possesses
generally two non-trivial flat directions too.
One of them exists only if $\tilde{\mu}^2<0$ and
lies at
\beq
\phi=\pm\,\sqrt{-\tilde{\mu}^2},\quad
\pb=\frac{2\ka\phi}{m}\,S,\quad
H^c=\Hb^c=0.
\label{semishift}
\eeq
It is a shifted flat direction with
\mbox{$V=\ka^2(M^4-\tilde{\mu}^4)$} along which
$G_{\rm PS}$ is broken to $G_{\rm SM}\times
{\rm U}(1)_{B-L}$. The second non-trivial
flat direction, which appears at
\begin{gather}
\phi=-\frac{\gamma m}{2\ka\la},\quad
\pb=-\frac{\gamma}{\la}\,S,\\
H^c\Hb^c=\frac{\ka\gamma(M^2-\phi^2)+\la m\phi}
{\gamma^2+\la^2},\\
V=V^0_\text{nsh}\equiv\frac{\ka^2\la^2}
{\gamma^2+\la^2}\left(M^2+\frac{\gamma^2m^2}
{4\ka^2\la^2}\right)^2,
\end{gather}
exists only for $\gamma\neq 0$ and is analogous
to the trajectory for the new shifted hybrid
inflation of Ref.~\cite{nshift}. Along this
direction, $G_{\rm PS}$ is broken to
$G_{\rm SM}$. In our subsequent discussion, we
will concentrate on the case $\tilde{\mu}^2>0$,
where the shifted flat direction in
Eq.~(\ref{semishift}) does not exist. It is
interesting to point out that, in this case, we
always have $V^0_\text{nsh}>V^0_{\rm tr}$ and it
is, thus, more likely that the system will
eventually settle down on the trivial rather
than the new shifted flat direction.

\par
If we expand the complex scalar fields $\phi$,
$\pb$, $H^c$, $\Hb^c$ in real and imaginary parts
according to the prescription $s=(s_1+i\,
s_2)/\sqrt{2}$, we find that, on the trivial flat
direction, the mass-squared matrices
$M_{\phi1}^2$ of $\phi_1$, $\pb_1$ and
$M_{\phi2}^2$ of $\phi_2$, $\pb_2$ are
\beq\label{eq:M2phi}
M_{\phi1(\phi2)}^2=\left(\ba{cc}
m^2+4\ka^2|S|^2\mp2\ka^2M^2 & -2\ka m S \\
-2\ka m S  & m^2 \ea\right)
\eeq
and the mass-squared matrices $M_{H1}^2$ of
$H^c_1$, $\bar{H}^c_1$ and $M_{H2}^2$ of $H^c_2$,
$\bar{H}^c_2$
are
\beq\label{eq:M2h}
M_{H1(H2)}^{2}=\left(\ba{cc}
\gamma^2|S|^2  & \mp\gamma\ka M^2  \\
\mp\gamma\ka M^2 & \gamma^2|S|^2 \ea\right).
\eeq
The matrices $M_{\phi1(\phi2)}^2$ are always
positive definite, while the matrices
$M_{H1(H2)}^2$ acquire one negative eigenvalue
for $|S|<S_c$. Thus, the trivial flat direction
is stable for $|S|>S_c$ and unstable for
$|S|<S_c$.

\par
It has been shown in Ref.~\cite{nsmooth} that,
for small enough values of the parameter
$\gamma$, the trivial flat direction, after
its destabilization at the critical point,
gives its place to a valley of {\em absolute}
minima for fixed $|S|$ which correspond to
$\theta\simeq 0$ and lead to the SUSY vacua in
Eq.~\eqref{eq:vacuum1}. This valley possesses an
inclination already at the classical level and
can accommodate a stage of inflation with the
properties of smooth hybrid inflation. In
Ref.~\cite{nsmooth}, the name new smooth hybrid
inflation was coined for the inflationary
scenario obtained when all the e-foldings required
for solving the horizon and flatness problems of
standard hot big bang cosmology are obtained when
the system follows this valley. In this paper, we
will study the case when the total required number
of e-foldings splits between two stages of
inflation, the standard hybrid inflation stage
for $|S|>S_c$ and the new smooth hybrid
inflation stage including an intermediate
inflationary period for $|S|<S_c$.

\par
The general outline of this scenario, which we
call standard-smooth hybrid inflation,
goes as follows. We assume that the system,
possibly after a period of pre-inflation at the
Planck scale, settles down on a point of the
trivial flat direction with $|S|>S_c$ (see e.g.
Ref.~\cite{init}). The constant classical
potential energy density on this direction breaks
SUSY explicitly and implies the existence of
one-loop radiative corrections which lift the
flatness of the potential producing the necessary
inclination for driving the inflaton towards the
critical point at $|S|=S_c$. So the standard
hybrid inflation stage of the scenario can be
realized along this path. As the system moves
below the critical point, some of the masses
squared of the fields become negative, resulting
to a phase of spinodal decomposition. This phase
is relatively fast, causes the spontaneous
breaking of $G_{\rm PS}$ to $G_{\rm SM}$, and
generates a limited number of e-foldings. After
this intermediate inflationary phase, the
system settles down on the new smooth hybrid
inflationary path and, thus, new smooth hybrid
inflation takes place. The second stage of
inflation consisting of the intermediate phase
and the subsequent new smooth hybrid inflation
yields the additional number of e-foldings
required for solving the horizon and flatness
problems of standard hot big bang cosmology. At
the end of this stage, the system falls rapidly
into the appropriate SUSY vacuum of the theory
leading, though, to no topological defect
production, since the GUT gauge group is already
broken to the SM gauge group during this
inflationary stage. Two more requirements need to
be fulfilled in order for this scenario to be
viable. First, one has to make sure that the
number of e-foldings generated during the
second stage of inflation is adequate for
diluting any monopoles generated during the
phase transition at the end of the first stage of
inflation. Secondly, one must ensure that all the
cosmologically relevant scales receive
inflationary perturbations only from the first
stage of inflation so that the existence of
measurable perturbations originating from the
phase of spinodal decomposition, which are of a
rather obscure nature, is avoided. Both of these
requirements are very easily satisfied in our
model, as we will see in the course of the
subsequent discussion.

\par
The one-loop radiative correction to the
potential due to the SUSY breaking on the trivial
inflationary path is calculated by the
Coleman-Weinberg formula \cite{ColemanWeinberg}:
\beq
\Delta V=\frac{1}{64\pi^2}\;\sum_i(-1)^{F_i}M_i^4
\ln\frac{M_i^2}{\Lambda^2},
\eeq
where the sum extends over all helicity states
$i$, $F_i$ and $M_i^2$ are the fermion number
and mass squared of the $i$th state and $\Lambda$
is a renormalization mass scale. In order to use
this formula for creating a logarithmic slope in
the inflationary potential, one has first to
derive the mass spectrum of the model on the
trivial inflationary path. It is easy to see
that, in the bosonic sector, one obtains two
groups of 45 pairs of real scalars with the
mass-squared matrices
\beq
M_{-(+)}^2=\left(\ba{cc}
m^2+4\ka^2|S|^2\mp2\ka^2M^2 & -2\ka m S \\
-2\ka m S  & m^2 \ea\right)
\eeq
and two more groups of 8 pairs of real scalars
with mass-squared matrices
\beq
M_{1(2)}^{2}=\left(\ba{cc}
\gamma^2|S|^2  & \mp\gamma\ka M^2  \\
\mp\gamma\ka M^2 & \gamma^2|S|^2 \ea\right).
\eeq
Note that $M_{-(+)}^2$ equals
$M_{\phi1(\phi2)}^2$ of Eq.~\eqref{eq:M2phi} and
$M_{1(2)}^2$ equals $M_{H1(H2)}^{2}$ of
Eq.~\eqref{eq:M2h}. In the fermionic sector of
the theory, we obtain 45 pairs of Weyl fermions
with mass-squared matrix
\beq
M_0^2=\left(\ba{cc}
m^2+4\ka^2|S|^2 & -2\ka m S \\
-2\ka m S  & m^2 \ea\right)
\eeq
and 8 more pairs of Weyl fermions with
mass-squared matrix
\beq
\bar{M}_0^2=\left(\ba{cc}
\gamma^2|S|^2  & 0  \\
0 & \gamma^2|S|^2 \ea\right).
\eeq
The matrices $M_0^2$, $\bar{M}_0^2$ equal
$M_{-(+)}^2$, $M_{1(2)}^{2}$ respectively
without the $\mp$ terms in the latter matrices.
The one-loop radiative correction to the
inflationary potential then takes the form
\bea
\label{eq:RadCorr}
\Delta V &=&
\frac{45}{64\pi^2}\;\tr\Big\{
M_{+}^4\ln\frac{M_{+}^2}{\Lambda^2}+
M_{-}^4\ln\frac{M_{-}^2}{\Lambda^2}
\nonumber\\
& &-2M_{0}^4\ln\frac{M_{0}^2}{\Lambda^2}
\Big\}
+\frac{8}{64\pi^2}\;\tr\Big\{
M_{1}^4\ln\frac{M_{1}^2}{\Lambda^2}
\nonumber\\
& &+
M_{2}^4\ln\frac{M_{2}^2}{\Lambda^2}-
2\bar{M}_{0}^4\ln\frac{\bar{M}_{0}^2}
{\Lambda^2}
\Big\}.
\eea
The total effective potential on the trivial
inflationary path will be given by
$V_{\rm tr}=v_0^4+\Delta V$, where
$v_0\equiv\sqrt{\kappa}M$ is the inflationary
scale. As already mentioned, the one-loop
radiative correction to the inflationary
potential lifts its classical flatness and
generates a logarithmic slope which is necessary
for driving the system towards the critical point
at $|S|=S_c$. It is important to note that the
$\sum_i(-1)^{F_i}M_i^4=8v_0^4\,(45\ka^2+
4\gamma^2)$ is $S$-independent, which implies
that the slope is $\Lambda$-independent and the
scale $\Lambda$, which remains undetermined, does
not enter the inflationary observables.

\par
Making the complex scalar field $S$ real by an
appropriate global ${\rm U}(1)$ R transformation
and defining the canonically normalized real
inflaton field $\si\equiv\sqrt{2}S$, the
slow-roll parameters $\veps$, $\eta$ and the
parameter $\xi^2$, which enters the running of
the spectral index, are (see e.g.
Ref.~\cite{review})
\bea
\veps &\equiv& \frac{m_{\rm P}^2}{2}\,
\left(\frac{V^\prime(\si)}{V(\si)}\right)^2,\\
\eta &\equiv& m_{\rm P}^2\,\left(
\frac{V^{\prime\prime}(\si)}{V(\si)}\right),\\
\xi^2 &\equiv& m_{\rm P}^4
\left(\frac{V^\prime(\si)V^{\prime\prime\prime}
(\si)}{V^2(\si)}\right),
\eea
where the prime denotes derivation with respect
to the inflaton $\si$ and $m_{\rm P}\simeq 2.44
\ten{18}\units{GeV}$ is the reduced Planck
mass. In these equations, $V$ is either the
effective potential $V_{\rm tr}$ on the trivial
inflationary path defined above, if we are
referring to the standard hybrid stage of
inflation, or the effective potential
$V_{\rm nsm}$ for new smooth hybrid inflation,
which has to be calculated numerically (see
Ref.~\cite{nsmooth}), if we are referring to the
new smooth hybrid inflationary phase.

\par
Numerical simulations have shown that, after
crossing the critical point at
$\si=\si_c\equiv\sqrt{2}S_c$, the system
continues evolving, for a while, with the Hubble
parameter $H$ remaining approximately constant
and equal to $H_0\equiv v_0^2/\sqrt{3}m_{\rm P}$
until it settles down on the new smooth hybrid
inflationary path at
$\si\approx0.99\,\si_c$. The scale factor of the
universe increases by about 8 e-foldings during
this intermediate period. The fields $H^c$ and
$\Hb^c$ are effectively massless at $\si=\si_c$
and, thus, acquire inflationary perturbations
$\delta H^c=\delta\Hb^c\approx H_0/2\pi$. Their
initial values at the critical point are taken
equal to these perturbations. The inflaton $\si$
is assumed to have an initial velocity given by
the slow-roll equation
\beq
\dot{\si}=-\frac{V_{\rm tr}^\prime(\si_c)}{3H_0},
\eeq
where the overdot denotes derivation with respect
to the cosmic time $t$ and the inclination
$V_{\rm tr}^\prime(\si_c)$ is provided by the
radiative corrections on the trivial flat
direction (for the parameter values that are
of interest, the slow-roll conditions
$\veps\leq 1$, $|\eta|\leq 1$ for the first stage
of inflation are violated only
``infinitesimally'' close to the critical point).
Although the above results are not independent
from the values of the model parameters, they
represent legitimate mean values. Moreover,
inflationary observables like the spectral index
have shown not to depend significantly on the
properties of this intermediate phase.

\par
From the above discussion, we see that the number
of e-foldings from the time when the
pivot scale $k_0=0.002\units{Mpc}^{-1}$ crosses
outside the inflationary horizon until the end of
inflation is (see e.g. Ref.~\cite{review})
\bea
\label{eq:NQ}
N_Q&\approx&\frac{1}{m_{\rm P}^2}\,
\int_{\si_f}^{0.99\,\si_c}\frac{V_{\rm nsm}(\si)}
{V_{\rm nsm}^\prime(\si)}\,d\si+8
\nonumber\\
& &+\frac{1}{m_{\rm P}^2}\,
\int_{\si_c}^{\si_Q}\frac{V_{\rm tr}(\si)}
{V_{\rm tr}^\prime(\si)}\,d\si,
\eea
where $\si_Q\equiv\sqrt{2} S_Q>0$ is the value
of the inflaton field at horizon crossing of the
pivot scale and $\si_f$ refers to the value of
$\si$ at the end of the second stage of inflation
and can be found from the corresponding slow-roll
conditions. The power spectrum $P_{\mathcal R}$
of the primordial curvature perturbation at the
scale $k_0$ is given (see e.g.
Ref.~\cite{review}) by
\beq\label{eq:Perturbations}
P_{\mathcal R}^{1/2}\simeq
\frac{1}{2\pi\sqrt{3}}\,\frac{V_{\rm tr}^{3/2}
(\si_Q)}{m_{\rm P}^3V_{\rm tr}^\prime(\si_Q)}.
\eeq
The spectral index $n_{\rm s}$, the
tensor-to-scalar
ratio $r$, and the running of the spectral index
$dn_{\rm s}/d\ln k$ can be written (see e.g.
Ref.~\cite{review}) as
\begin{gather}
n_{\rm s}\simeq 1+2\eta-6\veps,\quad
r\simeq\,16\veps,\nonumber\\
\frac{dn_{\rm s}}{d\ln k}\simeq16\veps\eta
-24\veps^2-2\xi^2,
\end{gather}
where $\veps$, $\eta$, and $\xi^2$ are evaluated
at $\si=\si_Q$. The number of e-foldings $N_Q$
required for solving the horizon and flatness
problems of standard hot big bang cosmology is
given (see e.g. Ref.~\cite{lectures})
approximately by
\beq\label{eq:NQvsVinf}
N_Q\simeq53.76\,+\frac{2}{3}\,\ln\left(\frac{v_0}
{10^{15}\units{GeV}}\right)+\frac{1}{3}\,\ln
\left(\frac{T_{\rm r}}{10^9\units{GeV}}\right),
\eeq
where $T_{\rm r}$ is the reheat temperature that
is expected not to exceed about
$10^9\units{GeV}$, which is the well-known
gravitino bound \cite{gravitino}.

\par
As already explained, magnetic monopoles are
produced at the end of the standard hybrid stage
of inflation, where $G_{\rm PS}$ breaks down
to $G_{\rm SM}$. We will now discuss, in some
detail, this production of magnetic monopoles and
their dilution by the subsequent second stage of
inflation. The masses of the fields $H^c$ and
$\Hb^c$, which vanish at $\si=\si_c$, grow very
fast as the system moves to smaller values of
$\si$. Actually, as one can show numerically,
they become of order $H_0$ when the system is
still ``infinitesimally'' close to the critical
point and the inflationary perturbations of $H^c$
and $\Hb^c$ become suppressed. After this, the
system evolves essentially classically. It
remains, for a while, close to the trivial flat
direction (which, for $\si<\si_c$, is unstable
as it consists of saddle points) yielding about 8
e-foldings as mentioned above. It, finally,
settles down on the new smooth hybrid
inflationary path at $\si\approx 0.99\,\si_c$.
To be more precise, it ends up at a point of the
manifold which consists of the {\em absolute}
minima of the
potential for fixed $\si\approx 0.99\,\si_c$.
The particular choice of this point is made by
the inflationary perturbations of $H^c$ and
$\Hb^c$, which cease to operate when the masses
of these fields reach the value $H_0$. This
happens after crossing the critical point, but
``infinitesimally'' close to it, as we already
mentioned. So the correlation length which is
relevant for magnetic monopole production by the
Kibble mechanism \cite{kibble} is
$\approx H_0^{-1}$.

\par
The initial monopole number density can then be
estimated \cite{kibble} as
\beq
n_{\rm M}^{\rm init}\approx\frac{3{\sf p}}{4\pi}
H_0^3,
\label{eq:inmondens}
\eeq
where ${\sf p}\sim 1/10$ is a geometric factor.
At the end of inflation, the monopole number
density becomes
\beq
n_{\rm M}^{\rm fin}\approx\frac{3{\sf p}}{4\pi}
H_0^3e^{-3\delta N},
\label{eq:finmondens}
\eeq
where $\delta N$ is the total number of
e-foldings during the intermediate period and
the subsequent new smooth hybrid inflation phase.
Dividing $n_{\rm M}^{\rm fin}$ by the number
density
$n_{\rm infl}\approx V^0_{\rm tr}/m_{\rm infl}$
of the inflatons which are produced at the
termination of inflation ($m_{\rm infl}$ is the
inflaton mass), we obtain that, at the end of
inflation, the number density of monopoles
$n_{\rm M}$ is given by
\beq
\frac{n_{\rm M}}{n_{\rm infl}}\approx
\frac{3{\sf p}}{4\pi}H_0^3e^{-3\delta N}
\frac{m_{\rm infl}}{V^0_{\rm tr}}.
\label{eq:monoverinfl}
\eeq
This ratio remains practically constant until
reheating, where the relative number density of
monopoles can be estimated as (compare with
Ref.~\cite{thermal})
\beq
\frac{n_{\rm M}}{\sf s}=\frac{n_{\rm M}}
{n_{\rm infl}}\frac{n_{\rm infl}}{\sf s}\approx
\frac{3{\sf p}}{16\pi}\frac{H_0T_{\rm r}}
{m_{\rm P}^2}e^{-3\delta N},
\label{eq:relmon}
\eeq
where ${\sf s}$ is the entropy density and the
relations $n_{\rm infl}/{\sf s}=3T_{\rm r}/4
m_{\rm infl}$ (in the instantaneous inflaton
decay approximation) and
$3H_0^2=V^0_{\rm tr}/m_{\rm P}^2$ were used.
After reheating, the relative number density of
monopoles remains essentially unaltered provided
that there is no entropy production at subsequent
times. Taking $n_{\rm M}/{\sf s}\lesssim
10^{-30}$, which corresponds \cite{monreldens} to
the Parker bound \cite{parker} on the present
magnetic monopole flux in our galaxy derived from
galactic magnetic field considerations,
$T_{\rm r}\sim 10^9\units{GeV}$, and $H_0\sim
10^{12}\units{GeV}$, we obtain from
Eq.~(\ref{eq:relmon}) that $\delta N\gtrsim 9.2$.
Using Eq.~\eqref{eq:NQvsVinf}, this implies that
$N_{\rm st}\lsim 45$, where $N_{\rm st}$ is the
number of e-foldings of the pivot scale $k_0$
during the standard hybrid stage of inflation.
Saturating this bound,
we obtain a monopole flux which may be
measurable. However, the interesting values of
$N_{\rm st}$ encountered here in the global SUSY
case are much smaller (see below) and, thus, the
predicted magnetic monopole flux is unlikely to
be measurable. In the minimal SUGRA case,
$N_{\rm st}$ is restricted to quite small values
(see Sec.~\ref{sec:sugra}) and the monopole flux
is predicted utterly negligible.

\par
The model contains five free parameters, namely
$M$, $m$, $\ka$, $\gamma$, and $\la$. As already
mentioned, the VEVs of $H^c$, $\bar{H}^c$ break
the PS gauge group to $G_{\rm SM}$, whereas the
VEV of the field $\phi$ breaks it only to
$G_{\rm SM}\times {\rm U}(1)_{B-L}$. So, the
gauge boson $A^\perp$ corresponding to the linear
combination of ${\rm U}(1)_{Y}$ and
${\rm U}(1)_{B-L}$ which is perpendicular to
${\rm U}(1)_{Y}$ acquires its mass squared
$m^2_{A^\perp}=(5/2)g^2|\vev{H^c}|^2$ solely from
the VEVs $\vev{H^c}$, $\vev{\bar{H}^c}$ of $H^c$,
$\bar{H}^c$ ($g$ is the SUSY
GUT gauge coupling constant). On the other hand,
the masses squared $m_A^2$ and $m_{W_{\rm R}}^2$
of the color triplet, anti-triplet ($A^\pm$) and
charged ${\rm SU}(2)_{\rm R}$ ($W^\pm_{\rm R}$)
gauge bosons get contributions from $\vev{\phi}$
too. Namely, $m_A^2=g^2(|\vev{H^c}|^2+(4/3)
|\vev{\phi}|^2)$ and $m_{W_{\rm R}}^2=g^2
(|\vev{H^c}|^2+2|\vev{\phi}|^2)$. As we will see
below, the VEVs of $H^c$ and $\phi$ in the SUSY
vacua of the model turn out to be of the same
order of magnitude. Since the $A^\pm$ gauge
bosons are expected to affect the renormalization
group equations to a greater extent than the
$W^\pm_{\rm R}$ ones (the SM singlet gauge boson
$A^\perp$ does not affect them at all), we set
the mass $m_A$ divided by $g\approx 0.7$ equal to
the SUSY GUT scale $M_{\rm GUT}$. We also set the
value of the parameter $p\equiv \sqrt{2}\ka M/m$
equal to $1/\sqrt{2}$. Note that, for
$\tilde{\mu}^2>0$, this parameter is smaller than
unity as seen from Eq.~(\ref{eq:mu2}). Finally,
we take $T_{\rm r}$ to saturate the gravitino
bound \cite{gravitino}, i.e. $T_{\rm r}\simeq
10^9\units{GeV}$, and fix the power
spectrum of the primordial curvature perturbation
to the WMAP3 \cite{wmap3} normalization
$P_{\mathcal R}^{1/2}\simeq 4.85\ten{-5}$ at the
pivot scale $k_0$. These choices fix three of the
five parameters of the model. So, we are left
with two free parameters. We will take the ratio
$\alpha\equiv|\vev{H^c}|/|\vev{\phi}|$, which,
for $\gamma$ adequately small, approximately
equals $\sqrt{m/\la M}$, to be one of them. The
second free parameter can be chosen to be the
number of e-foldings $N_{\rm st}$ of the pivot
scale $k_0$ during the standard hybrid stage of
inflation ($N_{\rm st}$ can be fixed by adjusting
e.g. the parameter $\gamma$). We will plot our
results as functions of these two free parameters.

\par
\begin{figure}[tp]
\centering
\includegraphics[width=\linewidth]{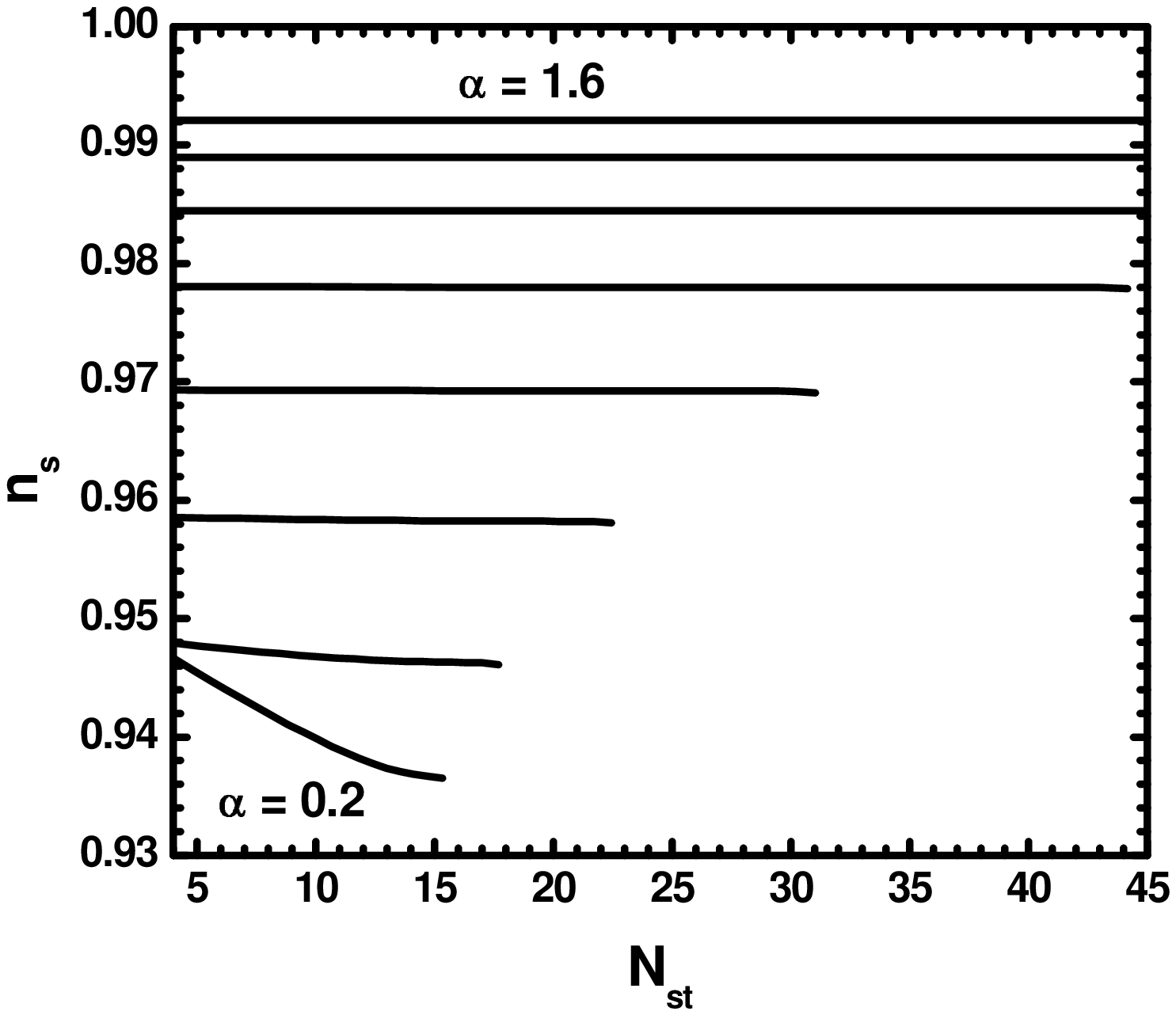}
\caption{Spectral index in standard-smooth hybrid
inflation versus $N_{\rm st}$ in global SUSY for
$p=\sqrt{2}\ka M/m=1/\sqrt{2}$. The values of the
parameter $\alpha=\vev{H^c}|/|\vev{\phi}|$ range
from $0.2$ to $1.6$ with steps of $0.2$.}
\label{fig:nsSUSY}
\end{figure}

In Fig.~\ref{fig:nsSUSY}, we plot the predicted
spectral index of the model versus the number of
e-foldings $N_{\rm st}$ suffered by the pivot
scale $k_0$ during the standard hybrid stage of
inflation for various values of the parameter
$\alpha$. Note that $N_{\rm st}$ is given by the
last term in the right-hand side of
Eq.~(\ref{eq:NQ}). We have restricted ourselves
to $N_{\rm st}$'s between 4 and 45. The lower
limit guarantees the validity of our requirement
that all the cosmological scales receive
perturbations from the first stage of inflation.
Indeed, the number of e-foldings that elapse
between the horizon crossing of the pivot scale
$k_0$ and the largest cosmological scale
$0.1/\units{Mpc}$ is about 4. The upper limit on
$N_{\rm st}$ ensures that the present flux of
magnetic monopoles in our galaxy does not exceed
the Parker bound \cite{parker} as we showed
above. The parameter $\alpha$ is limited between
0.2 and 1.6. Values of $\alpha$ lower than about
0.2 require non-perturbative values of $\lambda$,
whereas $\alpha=1.6$ or higher is of no much
interest since it leads to unacceptably large
$n_{\rm s}$'s. Whenever a curve in
Fig.~\ref{fig:nsSUSY} terminates on the right,
this means that the constraint
$P_{\mathcal R}^{1/2}\simeq 4.85 \ten{-5}$
cannot be satisfied beyond this endpoint. The
WMAP3 data fitted by the standard power-law
$\Lambda$CDM cosmological model predict
\cite{wmap3} that, at the pivot scale $k_0$,
\beq
\label{eq:nswmap3}
n_{\rm s}=0.958\pm 0.016~\Rightarrow~0.926
\lesssim n_{\rm s} \lesssim 0.99
\eeq
at $95\%$ confidence level. We see, from
Fig.~\ref{fig:nsSUSY}, that one can readily
obtain from our model spectral indices which lie
within this 2-$\si$ allowed range. Moreover, the
1-$\si$ range is fully covered by the predicted
values of $n_{\rm s}$. Note, however, that one
cannot obtain spectral indices lower than about
$0.936$. It is obvious that large values of
$N_{\rm st}$ are of no much interest since they
yield large $n_{\rm s}$'s. So a possibly
measurable flux of monopoles at the level of the
Parker bound is very unlikely.

\par
For the curves depicted in Fig.~\ref{fig:nsSUSY},
$\gamma$ varies in the range
$\gamma\simeq(0.04-6)\ten{-3}$. It increases
as $\alpha$ decreases or $N_{\rm st}$
increases with its dependence on $N_{\rm st}$
being much milder. The ranges of the
other parameters of the model are
$\ka\simeq(0.46-3.62)\ten{-2}$,
$\la\simeq 0.004-1.56$,
$M\simeq(1.45-2.44)\ten{16}\units{GeV}$,
$m\simeq(0.13-1.56)\ten{15}\units{GeV}$,
$\si_Q\simeq(0.9-8.8)\ten{17}\units{GeV}$,
$\si_c\simeq(0.8-2.3)\ten{17}\units{GeV}$,
and $\si_f\simeq(0.5-1.5)\ten{17}\units{GeV}$.
The total number of e-foldings
from the time when the pivot scale $k_0$
crosses outside the inflationary horizon until
the end of the second stage of inflation is
$N_Q\simeq 53.7-54.7$. Finally,
$dn_{\rm s}/d\ln k\simeq-(0.06-4)\ten{-3}$ and
the tensor-to-scalar ratio
$r\simeq(0.008-2.8)\ten{-4}$. A decrease in the
value of $p$, which is the only arbitrarily
chosen parameter, generally leads to an increase
of the spectral index. Thus, smaller values of
$p$ are expected to shift the curves in
Fig.~\ref{fig:nsSUSY} upwards, but otherwise
do not change the qualitative features of the
model.

\section{Supergravity corrections}
\label{sec:sugra}

We now turn to the discussion of the SUGRA
corrections to the inflationary potentials of our
model. The F--term scalar potential in SUGRA is
given by
\beq
\label{eq:VSUGRA}
V=e^{K/m_{\rm P}^2}
\left[(F_i)^* K^{i^*j}F_j-3\,\frac{|W|^2}
{m_{\rm P}^2}\right],
\eeq
where $K$ is the K\"{a}hler potential,
$F_i=W_i+K_iW/m_{\rm P}^2$, a subscript $i$
($i^*$) denotes derivation with respect to the
complex scalar field $s^i$ ($s^{i{\,*}}$) and
$K^{i^*j}$ is the inverse of the K\"{a}hler
metric $K_{j\,i^*}$. We will only consider
SUGRA with minimal K\"{a}hler potential and
show that the WMAP3 results \cite{wmap3} can be
met for a wide range of values of the parameters
of the model.

\par
The minimal K\"{a}hler potential in the model
under consideration has the form
\beq\label{eq:MinKahler}
K^{\rm min}=|S|^2+|\phi|^2+|\pb|^2+|H^c|^2+
|\Hb^c|^2
\eeq
and the corresponding F--term scalar potential is
\beq
\label{eq:SUGRAmin}
V^{\rm min}=e^{K^{\rm min}/m_{\rm P}^2}\;\left[
\sum_{s}\left|W_s+\frac{Ws^*}{m_{\rm P}^2}
\right|^2-3\,\frac{|W|^2}{m_{\rm P}^2}\right],
\eeq
where $s$ stands for any of the five complex
scalar fields appearing in
Eq.~\eqref{eq:MinKahler}. It is very easily
verified that, on the trivial flat direction,
this scalar potential expanded up to fourth
order in $|S|$ takes the form
\beq
V^{\rm min}_{\rm tr}\simeq v_0^4\left
(1+\frac{1}{2}\,\frac{|S|^4}{m_{\rm P}^4}\right).
\eeq
Thus, after including the SUGRA corrections with
minimal K\"{a}hler potential, the effective
potential during the standard hybrid stage of
inflation becomes
\beq
\label{Vtrsugra}
V^{\rm SUGRA}_{\rm tr}\simeq V^{\rm min}_{\rm tr}
+\Delta V
\eeq
with $\Delta V$ representing the one-loop
radiative correction given in
Eq.~\eqref{eq:RadCorr}. Furthermore, it has been
shown in Ref.~\cite{nsmooth} that the effective
potential on the new smooth hybrid inflationary
path in the presence of minimal SUGRA takes the
form
\beq
V^{\rm SUGRA}_{\rm nsm}\simeq v_0^4\left(
\tilde{V}_{\rm nsm}
+\frac{1}{2}\,\frac{|S|^4}{m_{\rm P}^4}\right),
\eeq
where $\tilde{V}_{\rm nsm}\equiv
V_{\rm nsm}/v_0^4$ with $V_{\rm nsm}$ being the
effective potential on the new smooth hybrid
inflationary path in the case of global SUSY.
Note that, in the minimal SUGRA case, the
critical value of $\sigma$, where the trivial
flat direction becomes unstable, will be slightly
different from the critical value of $\sigma$ in
the global SUSY case.

\par
The cosmology of the model after including the
minimal SUGRA corrections follows
straightforwardly from that of the global SUSY
case if one replaces the inflationary effective
potentials of the latter by the ones derived
above and take into account some changes in the
intermediate phase between the two main
inflationary periods. Actually, one finds
numerically that, due to the larger inclination
of the inflationary path provided by the minimal
SUGRA corrections, the number of e-foldings
during the intermediate period of inflation is
reduced to about 2 or 3. Also, the value of $\si$
at which the system settles down on the new
smooth hybrid inflationary path decreases to
about $\si\approx 0.95\,\si_c$. Moreover, as it
turns out, the evolution of the system can be
very well approximated by the simplifying
assumption that, during the intermediate phase,
it follows the new smooth hybrid inflationary
path. Therefore, we remove the term 8 from the
right-hand side of Eq.~(\ref{eq:NQ}) and replace
the upper limit in the first integral by $\si_c$.

\par
We again set the mass $m_A$ of the color triplet,
anti-triplet gauge bosons divided by
$g\approx 0.7$ equal to the SUSY GUT scale
$M_{\rm GUT}$ and the value of the parameter
$p=\sqrt{2}\ka M/m$ equal to $1/\sqrt{2}$.
We also take $T_{\rm r}$ to saturate the
gravitino bound \cite{gravitino}, i.e. $T_{\rm r}
\simeq 10^9\units{GeV}$, and fix the power
spectrum of the primordial curvature perturbation
to the WMAP3 \cite{wmap3} normalization
$P_{\mathcal R}^{1/2}\simeq 4.85 \ten{-5}$ at the
pivot scale $k_0$. Finally, we will again plot
our results against the parameter
$\alpha=|\vev{H^c}|/|\vev{\phi}|$ and the
number of e-foldings $N_{\rm st}$ of the pivot
scale $k_0$ during the standard hybrid stage of
inflation ($N_{\rm st}$ can again be fixed by
adjusting e.g. the parameter $\gamma$).

\begin{figure}[tp]
\centering
\includegraphics[width=\linewidth]{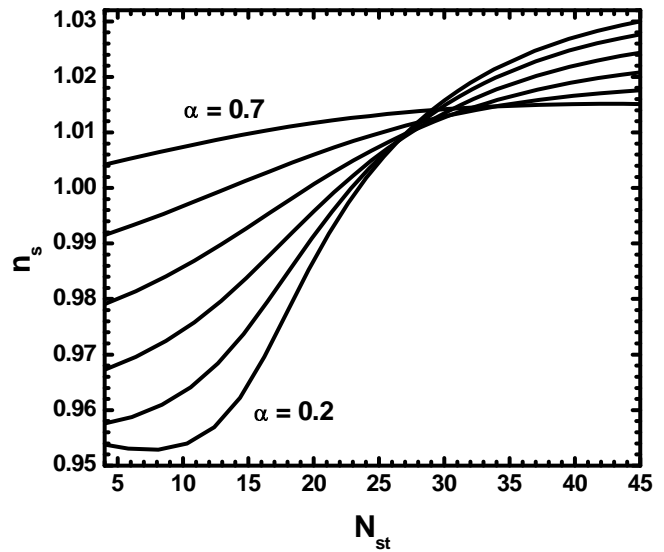}
\caption{Spectral index in standard-smooth hybrid
inflation versus $N_{\rm st}$ in minimal SUGRA
for $p=\sqrt{2}\ka M/m=1/\sqrt{2}$. The values of
the parameter $\alpha$ range from $0.2$ to $0.7$
with steps of $0.1$.}
\label{fig:nsSUGRA}
\end{figure}

\par
In Fig.~\ref{fig:nsSUGRA}, we plot the predicted
spectral index of the model in minimal SUGRA
versus $N_{\rm st}$ for various values of the
parameter $\alpha$. We have allowed
$N_{\rm st}$ to vary only between 4 and 45 for
the same reasons mentioned in the global SUSY
case. For $\alpha$ smaller than about 0.2, the
required values of $\lambda$ turn out again to be
non-perturbative, whereas, for $\alpha$ greater
than about 0.7, the WMAP3 normalization of the
power spectrum of the primordial curvature
perturbation is not satisfied. We see that
spectral indices below unity are readily
obtainable and that the central value $n_{\rm s}
=0.958$ from the WMAP3 results is achievable.
Though, the spectral index cannot be reduced below
$n_{\rm s}\simeq 0.953$, as is evident from the
curve with $\alpha=0.2$. Note that values of
$n_{\rm s}$ in the $95\%$ confidence level range
of Eq.~(\ref{eq:nswmap3}) can be obtain only if
$N_{\rm st}$ is lower than about 21. So, the
predicted magnetic monopole flux in our galaxy is
utterly negligible.

\par
The range of variance of the parameter $\gamma$
on the curves of Fig.~\ref{fig:nsSUGRA} is
$\gamma\simeq(0.17-3.43)\ten{-3}$ with $\gamma$
increasing with decreasing $\alpha$ and slightly
increasing with increasing $N_{\rm st}$. The
ranges of the other parameters of the model on
these curves are
$\ka\simeq(0.66-1.35)\ten{-2}$,
$\la\simeq 0.027-0.68$,
$M\simeq(2.12-2.44)\ten{16}\units{GeV}$,
$m\simeq(2.8-6.6)\ten{14}\units{GeV}$,
$\si_Q\simeq(0.95-3.05)\ten{17}\units{GeV}$,
$\si_c\simeq(0.6-2)\ten{17}\units{GeV}$,
and $\si_f\simeq(4.9-9.9)\ten{16}\units{GeV}$.
The total number of e-foldings from the time
when the pivot scale $k_0$ crosses outside the
inflationary horizon until the end of the second
stage of inflation is $N_Q\simeq 54.1-54.5$.
Finally, $dn_{rm s}/d\ln k\simeq-(0.77-3.76)
\ten{-3}$ and $r\simeq(0.7-5.3)\ten{-5}$. Again,
a decrease in the value of $p$ generally leads
to an increase of the spectral index, resulting,
thus, to a shift of the curves in
Fig.~\ref{fig:nsSUGRA} upwards. However, the
other qualitative features of the model are not
affected.

\section{Gauge unification}
\label{sec:gauge}

\par
We will now briefly address the question of gauge
unification in our model. As the careful reader
may have noticed, cosmological considerations
have constrained the mass parameter $m$ to be
significantly lower than $M_{\rm GUT}$,
especially in the case of minimal SUGRA. This
could easily jeopardize the unification of gauge
coupling constants and, indeed, it does, as it
turns out, since some of the fields that
contribute significantly to the gauge coupling
constant running acquire masses of order $m$.
Actually, there are two different scales below
$M_{\rm GUT}$ that give masses to fields
contributing to the renormalization group
equations for the gauge coupling constants. One
of them is, as already mentioned, around $m$ and
the other is around
$|\vev{H^c}|=\sqrt{m|\vev{\phi}|/\la}$. This
holds in the minimal SUGRA case and, for not too
large $n_{\rm s}$'s, in the global SUSY case
too. Gauge unification is destroyed for two
reasons. First of all, the fields which acquire
masses below $M_{\rm GUT}$ are too
many and this causes the appearance of Landau
poles in the running of the gauge coupling
constants. Secondly, none of these fields has
$\rm SU(2)_L$ quantum numbers and thus, even if
divergences were not present, the $\rm SU(2)_L$
gauge coupling constant would fail to unify with
the other gauge coupling constants.

\par
The first problem is avoided by considering the
superpotential term $\xi\phi^2\pb$, which
is allowed by all the symmetries of the theory
(see Ref.~\cite{quasi}). The reason for not
including this term in our discussion from the
beginning is that it does not contain a coupling
between the SM singlet components of $\phi$,
$\pb$ and so does not affect the inflationary
dynamics. This is because $\phi^2\pb$ is the
mixed product of the three vectors $\phi$,
$\phi$, and $\pb$ in the 3-dimensional space in
which the $\rm SO(3)$ group which is locally
equivalent to $\rm SU(2)_R$ operates.
Nevertheless, this term generates extra
contributions of order $|\xi\vev{\phi}|^2$ to
the masses squared of some fields and, thus,
helps us to get rid of the Landau poles.

\par
The second problem can be solved only by
including extra fields in the model which affect
the running of the $\rm SU(2)_L$ gauge
coupling constant. Note that, although the
extended PS model under consideration already
contains fields with $\rm SU(2)_L$ quantum
numbers which are not present in the minimal
SUSY PS model, namely the fields $h'$ and
$\bar{h}'$ belonging to the (15,2,2)
representation (see Ref.~\cite{quasi}), these
fields are not sufficient for achieving the
desired gauge unification since they do not
affect the running of the $\rm SU(2)_L$ gauge
coupling constant as much as it is required.
Consequently, one has to consider the inclusion
of some extra fields. There is a good choice
which uses a single extra field, namely a
superfield $\chi$ belonging to the (15,3,1)
representation. If we require that this field has
charge $1/2$ under the global $\rm U(1)$ R
symmetry, then the only superpotential term in
which this field is allowed to participate is a
mass term of the form
$\frac{1}{2}m_{\chi}\chi^2$. One can then tune
the new mass parameter $m_{\chi}$ so as to
achieve unification of the gauge coupling
constants. We find that this mass should be
$\approx 8\ten{14}\units{GeV}$.

\par
It turns out that one can achieve gauge
unification at the appropriate scale
($\approx 2\ten{16}\units{GeV}$) as long as the
mass parameter $m$ is constrained to lie above
$3\ten{14}\units{GeV}$. This condition is
fulfilled for almost all curves of
Figs.~\ref{fig:nsSUSY} and \ref{fig:nsSUGRA}
except for the curves with $\alpha=1.2$, $1.4$,
and $1.6$ in Fig.~\ref{fig:nsSUSY}. Note that
this constraint is equivalent to the statement
that the spectral index in the global SUSY case
is less than about $0.98$. So, the low spectral
index regime is not affected. Furthermore, if
one wants to be on the safe side avoiding
marginal gauge unification (the value
$m\approx 3\ten{14}\units{GeV}$ leads to gauge
unification with a rather large GUT gauge
coupling constant, which is of order unity or
larger), then one can impose the restriction
$m\gsim 4\ten{14}\units{GeV}$, which leads to
the constraints $\alpha\lsim 0.8$ for
Fig.~\ref{fig:nsSUSY} and $\alpha\lsim 0.5$ for
Fig.~\ref{fig:nsSUGRA}.

\section{Conclusions}
\label{sec:conclusions}

\par
We have reconsidered the extended SUSY PS model
of Ref.~\cite{quasi} which solves the $b$-quark
mass problem. In this model, exact asymptotic
Yukawa unification is naturally and moderately
violated so that, for $\mu>0$, the predicted
$b$-quark mass lies within the experimentally
allowed range even with universal boundary
conditions. The same model can automatically
lead to new versions of the shifted and smooth
hybrid inflationary scenarios based solely on
renormalizable superpotential interactions. In
both of these cases, the PS GUT gauge group is
broken to the SM gauge group already during
inflation and, thus, no PS magnetic monopole
production takes place at the end of inflation.
So, the possible cosmological catastrophe from
magnetic monopole overproduction is avoided.
In contrast to new smooth hybrid inflation,
the new shifted one yields, in global SUSY,
spectral indices which are too close to unity
and without much running in conflict with the
recent WMAP3 data. Moreover, inclusion of
minimal SUGRA raises $n_{\rm s}$ to unacceptably
large values in both of these inflationary
scenarios.

\par
To resolve this problem, we proposed a two-stage
inflationary scenario which is naturally realized
within this extended SUSY PS model for the range
of its parameters leading to new smooth hybrid
inflation. The first stage of inflation is of the
standard hybrid type and takes place along the
trivial classically flat direction of the scalar
potential, which is stable for values of the
inflaton field larger than a certain critical
value. The inflaton is driven by the logarithmic
slope acquired by this direction from one-loop
radiative corrections which are due to the SUSY
breaking caused by the non-vanishing potential
energy density on this direction. Note that, on
the trivial flat direction, the PS gauge group is
unbroken. Assuming that the cosmological scales
exit the horizon during the first stage of
inflation, we can achieve, in global SUSY,
spectral indices compatible with the WMAP3 data
by restricting the number of e-foldings suffered
by our present horizon scale during this
inflationary stage.

\par
The system, after crossing the critical point of
the trivial flat direction, undergoes a
relatively short intermediate inflationary phase
and then falls rapidly into the new smooth hybrid
inflationary path along which it continues
inflating as it slowly rolls towards the vacua.
Note that this path appears right after the
destabilization of the trivial flat direction at
its critical point. During this second stage of
(intermediate plus new smooth hybrid) inflation,
the additional number of e-foldings needed for
solving the horizon and flatness problems is
naturally generated and $G_{\rm PS}$ is broken to
$G_{\rm SM}$. So, we see that the necessary
complementary inflation is automatically built in
the model itself and we do not have to invoke an
{\it ad hoc} second stage of inflation as in
other scenarios. Moreover, large reheat
temperatures can be achieved after the second
stage of inflation since this stage is realized
at a superheavy scale. Therefore, baryogenesis
via (non-thermal) leptogenesis may work in
this case in contrast to other models where the
reheat temperature is too low for sphalerons to
operate. Finally, the PS monopoles that are
formed at the end of the standard hybrid stage
of inflation can be adequately diluted by the
second stage of inflation. The monopole flux in
our galaxy in the case of global SUSY is
expected to be utterly negligible for not too
large values of the spectral index.

\par
Including SUGRA corrections with minimal
K\"{a}hler potential enhances the predicted
values of the spectral index, which, however,
remain within the allowed interval for a wide
range of the model parameters. So, in this model,
there is no need to include non-minimal terms in
the K\"{a}hler potential and, thus, complications
from the possible appearance of a local maximum
and minimum of the inflationary potential are
avoided. The monopole flux in the SUGRA case
turns out not to be measurable for all the
allowed values of the model parameters.

\section*{ACKNOWLEDGEMENTS}
\par
We thank I.N.R. Peddie for his help with
gauge unification. This work was
supported by the European Union under the
contracts MRTN-CT-2004-503369 and
HPRN-CT-2006-035863 as well as the Greek
Ministry of Education and Religion and the
EPEAK program Pythagoras.

\def\ijmp#1#2#3{{Int. Jour. Mod. Phys.}
{\bf #1},~#3~(#2)}
\def\plb#1#2#3{{Phys. Lett. B }{\bf #1},~#3~(#2)}
\def\zpc#1#2#3{{Z. Phys. C }{\bf #1},~#3~(#2)}
\def\prl#1#2#3{{Phys. Rev. Lett.}
{\bf #1},~#3~(#2)}
\def\rmp#1#2#3{{Rev. Mod. Phys.}
{\bf #1},~#3~(#2)}
\def\prep#1#2#3{{Phys. Rept. }{\bf #1},~#3~(#2)}
\def\prd#1#2#3{{Phys. Rev. D }{\bf #1},~#3~(#2)}
\def\npb#1#2#3{{Nucl. Phys. }{\bf B#1},~#3~(#2)}
\def\npps#1#2#3{{Nucl. Phys. B (Proc. Sup.)}
{\bf #1},~#3~(#2)}
\def\mpl#1#2#3{{Mod. Phys. Lett.}
{\bf #1},~#3~(#2)}
\def\arnps#1#2#3{{Annu. Rev. Nucl. Part. Sci.}
{\bf #1},~#3~(#2)}
\def\sjnp#1#2#3{{Sov. J. Nucl. Phys.}
{\bf #1},~#3~(#2)}
\def\app#1#2#3{{Acta Phys. Polon.}
{\bf #1},~#3~(#2)}
\def\rnc#1#2#3{{Riv. Nuovo Cim.}
{\bf #1},~#3~(#2)}
\def\ap#1#2#3{{Ann. Phys. }{\bf #1},~#3~(#2)}
\def\ptp#1#2#3{{Prog. Theor. Phys.}
{\bf #1},~#3~(#2)}
\def\apjl#1#2#3{{Astrophys. J. Lett.}
{\bf #1},~#3~(#2)}
\def\n#1#2#3{{Nature }{\bf #1},~#3~(#2)}
\def\apj#1#2#3{{Astrophys. J.}
{\bf #1},~#3~(#2)}
\def\anj#1#2#3{{Astron. J. }{\bf #1},~#3~(#2)}
\def\apjs#1#2#3{{Astrophys. J. Suppl.}
{\bf #1},~#3~(#2)}
\def\mnras#1#2#3{{MNRAS }{\bf #1},~#3~(#2)}
\def\grg#1#2#3{{Gen. Rel. Grav.}
{\bf #1},~#3~(#2)}
\def\s#1#2#3{{Science }{\bf #1},~#3~(#2)}
\def\baas#1#2#3{{Bull. Am. Astron. Soc.}
{\bf #1},~#3~(#2)}
\def\ibid#1#2#3{{\it ibid. }{\bf #1},~#3~(#2)}
\def\cpc#1#2#3{{Comput. Phys. Commun.}
{\bf #1},~#3~(#2)}
\def\astp#1#2#3{{Astropart. Phys.}
{\bf #1},~#3~(#2)}
\def\epjc#1#2#3{{Eur. Phys. J. C}
{\bf #1},~#3~(#2)}
\def\nima#1#2#3{{Nucl. Instrum. Meth. A}
{\bf #1},~#3~(#2)}
\def\jhep#1#2#3{{J. High Energy Phys.}
{\bf #1},~#3~(#2)}
\def\lnp#1#2#3{{Lect. Notes Phys.}
{\bf #1},~#3~(#2)}
\def\appb#1#2#3{{Acta Phys. Polon. B}
{\bf #1},~#3~(#2)}
\def\njp#1#2#3{{New J. Phys.}
{\bf #1},~#3~(#2)}
\def\pl#1#2#3{{Phys. Lett. }{\bf #1B},~#3~(#2)}
\def\jcap#1#2#3{{J. Cosmol. Astropart. Phys.}
{\bf #1},~#3~(#2)}
\def\mpla#1#2#3{{Mod. Phys. Lett. A}
{\bf #1},~#3~(#2)}
\def\jetpsp#1#2#3{{JETP (Sov. Phys.)}
{\bf #1},~#3~(#2)}
\def\jetpl#1#2#3{{JETP Lett.}
{\bf #1},~#3~(#2)}
\def\jpa#1#2#3{{J. Phys. A}
{\bf #1},~#3~(#2)}
\def\apjss#1#2#3{{Astrophys. J. Suppl. Ser.}
{\bf #1},~#3~(#2)}


\begin{thebibliography}{99}

\bibitem{smooth1}
G. Lazarides and C. Panagiotakopoulos,
\prd{52}{1995}{R559}.

\bibitem{cop}
E.J. Copeland, A.R. Liddle, D.H. Lyth,
E.D. Stewart, and D. Wands,
\prd{49}{1994}{6410}.

\bibitem{rc}
G.R. Dvali, Q. Shafi, and R.K. Schaefer,
\prl{73}{1994}{1886};
G. Lazarides, R.K. Schaefer, and Q. Shafi,
\prd{56}{1997}{1324}.

\bibitem{linde}
A.D. Linde, \prd{49}{1994}{748}.

\bibitem{string}
H.B. Nielsen and P. Olesen,
\npb{61}{1973}{45};
G. Lazarides, Q. Shafi, and T.F. Walsh,
{\it ibid.} {\bf B195}, 157 (1982).

\bibitem{monopole}
G.'t Hooft, \npb{79}{1974}{276};
A.M. Polyakov, \jetpl{20}{1974}{194};
J.P. Preskill, \prl{43}{1979}{1365};
G. Lazarides, Q. Shafi, and W.P. Trower,
\ibid{49}{1982}{1756}.

\bibitem{domain}
Ya.B. Zeldovich, I.Yu. Kobzarev, and L.B. Okun,
\jetpsp{40}{1974}{1}.

\bibitem{pati}
J.C. Pati and A. Salam, \prd{10}{1974}{275}.

\bibitem{magg}
G. Lazarides, M. Magg, and Q. Shafi,
\plb{97}{1980}{87}.

\bibitem{shift}
R. Jeannerot, S. Khalil, G. Lazarides, and
Q. Shafi, \jhep{10}{2000}{012}.

\bibitem{smooth2}
G. Lazarides, C. Panagiotakopoulos, and
N.D. Vlachos, \prd{54}{1996}{1369};
R. Jeannerot, S. Khalil, and G. Lazarides,
\plb{506}{2001}{344}.

\bibitem{talks}
G. Lazarides, hep-ph/0011130;
R. Jeannerot, S. Khalil, and G. Lazarides,
hep-ph/0106035.

\bibitem{nshift}
R. Jeannerot, S. Khalil, and G. Lazarides,
\jhep{07}{2002}{069}.

\bibitem{nsmooth}
G. Lazarides and A. Vamvasakis,
Phys. Rev. D {\bf 76}, 083507 (2007).

\bibitem{quasi}
M.E. Gomez, G. Lazarides, and C. Pallis,
\npb{638}{2002}{165}.

\bibitem{quasitalks}
G. Lazarides and C. Pallis, hep-ph/0404266;
hep-ph/ 0406081.

\bibitem{als}
B. Ananthanarayan, G. Lazarides, and
Q. Shafi, \prd{44}{1991}{1613};
\plb{300}{1993}{245}.

\bibitem{hw}
G. Lazarides and C. Panagiotakopoulos,
\plb{337}{1994}{90};
S. Khalil, G. Lazarides, and C. Pallis,
\ibid{508}{2001}{327}.

\bibitem{hall}
R. Hempfling, \prd{49}{1994}{6168};
L.J. Hall, R. Rattazzi, and U. Sarid,
\ibid{50}{1994}{7048}.

\bibitem{wetterich}
G. Lazarides, Q. Shafi, and C. Wetterich,
\npb{181}{1981}{287};
G. Lazarides and Q. Shafi,
\ibid{B350}{1991}{179}.

\bibitem{wmap3}
D.N. Spergel {\it et al.}, \apjs{170}{2007}{377}.

\bibitem{senoguz}
V.N. \c{S}eno\u{g}uz and Q. Shafi,
\plb{567}{2003}{79};
\ibid{582}{2004}{6}.

\bibitem{lofti}
L. Boubekeur and D. Lyth, \jcap{07}{2005}{010}.

\bibitem{king}
M. Bastero-Gil, S.F. King, and Q. Shafi,
\plb{651}{2007}{345}.

\bibitem{rehman}
M. ur Rehman, V.N. \c{S}eno\u{g}uz, and Q. Shafi,
\prd{75}{2007}{043522}.

\bibitem{gpp}
B. Garbrecht, C. Pallis, and A. Pilaftsis,
\jhep{12}{2006}{038}.

\bibitem{mhin}
G. Lazarides and C. Pallis, \plb{651}{2007}{216};
G. Lazarides, arXiv:0706.1436.

\bibitem{modular}
P. Bin\'{e}truy and M.K. Gaillard, Phys. Rev. D
{\bf 34}, 3069 (1986);
F.C. Adams, J.R. Bond, K. Freese, J.A. Frieman,
and A.V. Olinto, \ibid{47}{1993}{426};
T. Banks, M. Berkooz, S.H. Shenker, G.W. Moore,
and P.J. Steinhardt, \ibid{52}{1995}{3548};
R. Brustein, S.P. De Alwis, and E.G. Novak,
\ibid{68}{2003}{023517}.

\bibitem{yamaguchi} M. Kawasaki, M. Yamaguchi,
and J. Yokoyama, \prd{68}{2003}{023508};
M. Yamaguchi and J. Yoko-yama,
\ibid{68}{2003}{123520};
\ibid{70}{2004}{023513};
M. Kawasaki, T. Takayama, M. Yamaguchi, and
J. Yokoyama, \ibid{74}{2006}{043525}.

\bibitem{thermallepto} M. Fukugita and
T. Yanagida,
Phys. Lett. B {\bf 174}, 45 (1986).

\bibitem{nonthermallepto} G. Lazarides and
Q. Shafi, \plb{258}{1991}{305};
G. Lazarides, C. Panagiotakopoulos, and Q. Shafi,
\ibid{315}{1993}{325}, (E) {\bf 317}, 661 (1993);
G. Lazarides, Q. Shafi, and N.D. Vlachos,
\ibid{427}{1998}{53};
G. Lazarides and N.D. Vlachos,
\ibid{459}{1999}{482}.

\bibitem{flat}
T. Gherghetta, C.F. Kolda, and S.P. Martin,
\npb{468}{1996}{37}.

\bibitem{quasithermal}
R. Allahverdi and A. Mazumdar,
\jcap{10}{2006}{008};
hep-ph/0603244.

\bibitem{mssminf}
R. Allahverdi, K. Enqvist, J. Garcia-Bellido, and
A. Mazumdar, \prl{97}{2006}{191304};
R. Allahverdi, B. Dutta, and A. Mazumdar,
\prd{75}{2007}{075018};
R. Allahverdi, K. Enqvist, J. Garcia-Bellido,
A. Jokinen, and A. Mazumdar,
\jcap{06}{2007}{019};
R. Allahverdi, A. Kusenko, and A. Mazumdar,
\ibid{07}{2007}{018};
R. Allahverdi, B. Dutta, and A. Mazumdar,
arXiv:0708.3983.

\bibitem{laz1}
G. Lazarides, C. Panagiotakopoulos, and Q. Shafi,
\prl{56}{1986}{432}.

\bibitem{laz2}
N. Ganoulis, G. Lazarides, and Q. Shafi,
\npb{323}{1989}{374}.

\bibitem{init}
G. Lazarides and N.D. Vlachos,
Phys. Rev. D {\bf 56}, 4562 (1997);
G. Lazarides and N. Tetradis,
{\it ibid.} {\bf 58}, 123502 (1998);
C. Panagiotakopoulos and N. Tetradis,
\ibid{59}{1999}{083502}.

\bibitem{ColemanWeinberg}
S.R. Coleman and E. Weinberg,
Phys. Rev. D {\bf 7}, 1888 (1973).

\bibitem{review}
B.A. Bassett, S. Tsujikawa, and D. Wands,
\rmp{78}{2006}{537}.

\bibitem{lectures}
G. Lazarides, \lnp{592}{2002}{351},
hep-ph/0111328;
hep-ph/0607032.

\bibitem{gravitino}
M.Yu. Khlopov and A.D. Linde,
Phys. Lett. B {\bf 138}, 265 (1984);
J. Ellis, J.E. Kim, and D. Nanopoulos,
\ibid{145}{1984}{181};
J.R. Ellis, D.V. Nanopoulos, and S. Sarkar,
\npb{259}{1985}{175}.

\bibitem{kibble} T.W.B. Kibble,
\jpa{9}{1976}{1387};
\prep{67}{1980}{183}.

\bibitem{thermal}
G. Lazarides and Q. Shafi, \plb{489}{2000}{194}.

\bibitem{monreldens} G. Lazarides,
C. Panagiotakopoulos, and Q. Shafi,
\prl{58}{1987}{1707}.

\bibitem{parker} E.N. Parker,
\apj{160}{1970}{383};
S.A. Bludman and M.A. Ruderman,
Phys. Rev. Lett. {\bf 36}, 840 (1976);
G. Lazarides, Q. Shafi, and T.F. Walsh,
\plb{100}{1981}{21}.

\end{thebibliography}
\end{document}